\documentclass[fleqn,a4paper,12pt]{article}
\usepackage{amssymb}
\usepackage{amsmath}
\usepackage[dvips]{graphicx}
\usepackage{placeins}
\usepackage{lscape}
\usepackage{a4}
\usepackage{epsfig}

\catcode`@=11 \long
\def\@caption#1[#2]#3{\par\addcontentsline{\csname
ext@#1\endcsname}{#1} {\protect\numberline{\csname
the#1\endcsname}{\ignorespaces #2}} \begingroup \small
\@parboxrestore \@makecaption{\csname fnum@#1\endcsname}
{\ignorespaces #3}\par \endgroup} \catcode`@=12

\renewcommand{\bar}{\overline}

\newcommand{\p}{\partial}
\newcommand{\vecphi}{\begin{pmatrix}
  \phi \\
  \xi
 \end{pmatrix}}

\newcommand{\C}{\mathbb{C}}
\newcommand{\R}{\mathbb{R}}

\newcommand{\N}{\mathbb{N}}

\newcommand{\diag}[2]{\mathrm{diag}\left\{#1,#2\right\}}
\newcommand{\abs}[1]{\left|#1\right|}

\DeclareMathOperator{\sech}{sech}
\DeclareMathOperator{\csch}{csch}

\begin{document}
\allowdisplaybreaks
\begin{titlepage} \vskip 2cm

\begin{center}
{\Large\bf Matrix Superpotential Linear in Variable Parameter}
\footnote{E-mail: {\tt yuri.karadzhov@gmail.com}}
\vskip 3cm {\bf {Yuri Karadzhov} \vskip 5pt
{\sl Institute of Mathematics, National Academy of Sciences of
Ukraine,\\ 3 Tereshchenkivs'ka Street, Kyiv-4, Ukraine, 01601\\}}
\end{center}
\vskip .5cm \rm
\begin{abstract}
The paper presents the classification of matrix valued superpotentials
corresponding to shape invariant systems of Schr\"odinger equations.
All inequivalent irreducible matrix superpotentials realized by matrices
of arbitrary dimension with linear dependence on variable parameter are presented explicitly.
\end{abstract}

\end{titlepage}

\section{Introduction \label{intro}}

Supersymmetric quantum mechanic presents a powerful and elegant tool for obtaining explicit solutions of
quantum mechanics problems described by Schr\"odinger equations~\cite{Witten}. Invented by Gendenstein \cite{Gen},
property of discrete reparametrization of potentials, known as shape-invariance, helps to determine whether
eigenvalues of Hamiltonians can be calculated by algebraic methods.

Though shape-invariant potentials
do not exhaust the full class of potentials of solvable Schr\"odinger equations \cite{hui}, it was interesting to find
new integrable models. Previously several attempts of describing the class of shape-invariant potentials were made.
In paper of Cooper et al \cite{Khare} a wide class of scalar shape-invariant potentials was described.

An attractive example of a matrix problem which admits a shape invariant supersymmetric formulation
was discovered by Pron'ko and Stroganov \cite{Pron}, who studied a motion of a neutral non-relativistic
fermion which interacts anomalously with the magnetic field generated by a thin current carrying wire.
The supersymmetric approach to the Pron'ko-Stroganov problem was applied in papers \cite{Vor}, \cite{Gol}, \cite{ninni}.

Particular cases of matrix potentials were discussed in papers \cite{Andr1}, \cite{Andr2}, \cite{Andri}, \cite{Rodr}.
Matrix superpotentials appear also in some supersymmetric systems
related to the crystalline structures in Gross-Neveu model \cite{pl1}, \cite{pl2}.
Two-dimensional matrix superpotentials, including shape invariant, were studied in papers \cite{tk1}, \cite{tk2}.
In paper~\cite{Fu} Fukui considers a certain class of shape invariant potentials,
which however was {\it ad hoc} restricted to $2\times2$ matrices with superpotentials linearly dependent
on the variable parameter.

Thus, in contrast to the class of scalar potentials, the class of known matrix potentials is presented by important
but rather particular examples. The remaining part of the class is still undiscovered and requires further research.
It seems to be interesting to widen the class of shape-invariant matrix potentials,
because in such a way we will be able to describe new exactly-integrable systems of Schr\"odinger equations.

A systematic study of the problem was carried out in our recent
paper \cite{yur1} where we presented a complete description of irreducible matrix
potentials, which include a term linear in variable parameter, and is proportional to the
unit matrix.

In the present paper the classification is continued and Fukui's results are generalized.
All shape-invariant matrix potentials of an arbitrary dimension with superpotentials
linearly dependent on the variable parameter were found.

\section{Shape-invariant potentials \label{sip}}

Let's start with a spectral problem
\begin{gather}
\label{specprob}
H_k\psi=E_k\psi,
\end{gather}
where $H_k$ is a Hamiltonian with a matrix potential, $E_k$ and $\psi$ are its
eigenvalues and eigenfunctions correspondingly, moreover $\psi$ is $n$-component spinor.
In the Schr\"odinger equation Hamiltonian has the form
\begin{gather}
\label{schrodeq}
H_k=-\frac{\p^2}{\p x^2}+V_k(x),
\end{gather}
where $V_k(x)$ is a $n$-dimensional matrix potential depending on variable $x$ and
parameter $k$.

Suppose that Hamiltonian accepts factorization
\begin{gather}
\label{hfact}
H_k=a_k^\dag a_k,
\end{gather}
then its superpartner is defined as follows
\begin{gather}
H_k^+ =a_ka_k^\dag .
\end{gather}
The most common representation of operators $a_k$ and $a_k^\dag$ has the form
\begin{gather}
\label{comrep}
a_k=A_k(x)\frac{\p}{\p x}+B_k(x),\quad a_k^\dag =-\frac{\p}{\p x}A_k^\dag (x)+B_k^\dag (x),
\end{gather}
where $A_k(x), B_k(x)$ are matrices depending on $x$ and $A_k^\dag (x), B_k^\dag (x)$
are hermitian conjugate to them.

Substituting this representation into equation (\ref{hfact}) we obtain the equation
\begin{gather}
H_k=-A_k^\dag A_k\frac{\p^2}{\p x^2}+(B_k^\dag A_k-A_k^\dag B_k-(A_k^\dag A_k)')\frac{\p}{\p x}+B_k^\dag B_k-(A_k^\dag B_k)',
\end{gather}
which is supposed to be Schr\"odinger equation of the form (\ref{schrodeq}). It leads to the following conditions
\begin{gather}
\begin{split} &
A_k^\dag A_k=I, \\ &
B_k^\dag A_k-A_k^\dag B_k-(A_k^\dag A_k)'=0, \\ &
B_k^\dag B_k-(A_k^\dag B_k)'=V_k.
\end{split}
\end{gather}
In terms of new variable $W_k(x)=A_k^\dag (x)B_k(x)$ this condition take the form
\begin{gather}
\begin{split} &
W_k^\dag =W_k, \\ &
V_k=W_k^2-W_k'.
\end{split}
\end{gather}

The same result can be obtained analogously with simpler representation of operators $a_k$ and $a_k^\dag$
\begin{gather}
\label{standrep}
a_k=\frac{\p}{\p x}+W_k(x),\quad a_k^\dag =-\frac{\p}{\p x}+W_k(x).
\end{gather}
$W_k(x)$ is called a matrix superpotential.

As $W_k(x)$ is hermitian, then the corresponding potential and its superpartner $V_k^+ (x)$, i.e.
\begin{gather}
V_k=-\frac{\p W_k}{\p x}+W_k^2,\quad V_k^+ =\frac{\p W_k}{\p x}+W_k^2
\end{gather}
are hermitian too.

The goal is to find such superpotentials which generate shape-invariant Hamiltonians
\begin{gather}
\label{hshapeinv}
H_k^+ =H_{F_k}+C_k,
\end{gather}
where $F_k, C_k$ are scalar functions of $k$. In terms of superpotential,
the last condition has the form
\begin{gather}
\label{shapeinv}
W_k^2+W'_k=W_{F_k}^2-W'_{F_k}+C_k.
\end{gather}

It is sufficient to search for irreducible superpotentials, which means matrix $W_k$
can't be transformed to block-diagonal form with a constant unitary transformation.

Note that if $F_k=k$ then equation (\ref{shapeinv}) transforms into
\begin{gather}
W_k'=\frac12C_k
\end{gather}
and it follows that
\begin{gather}
W_k=\frac12C_kx+\Omega_k,
\end{gather}
where $\Omega_k$ is hermitian matrix depending on $k$ which can be diagonalized. So in this
case superpotential is completely reducible to the one-dimensional shifted oscillators.

In case $F_k\neq k$ it makes sense to restrict ourselves to unit shifts
\begin{gather}
\label{shift}
F_k=k+1,
\end{gather}
see discussion section for details.

In the following sections the classification of superpotentials linearly dependent
on parameter $k$ is presented.

\section{The determining equations \label{deteq}}

Let us consider the superpotential of the form
\begin{gather}
\label{superpot}
W_k=kQ+P,
\end{gather}
where $P$ and $Q$ are $n\times n$ hermitian matrices dependent on $x$.

Suppose that $Q$ is not proportional to the unit matrix because in this case
the superpotential is reducible.

Substituting (\ref{superpot}) into (\ref{shapeinv}) and taking into account
equation (\ref{shift}) we obtain the equation
\begin{gather}
(2k+1)(Q'-Q^2) - \{Q,P\} + 2P'= C_k,
\end{gather}
where $Q'=\frac{\p Q}{\p x}, P'=\frac{\p P}{\p x}$ and $\{Q,P\}=QP+PQ$ is an anticommutator of matrices $Q$ and $P$.
After variable separation the last equation transforms into a system of determining equations
\begin{gather}
\label{qeq}
Q'=Q^2+\nu, \\
\label{peq}
P'=\frac12\{Q,P\}-\mu, \\
\label{ceq}
C_k=(2k+1)\nu - 2\mu,
\end{gather}
where $\mu, \nu \in\R$ are arbitrary constants.
For scalar values unit matrix $I$ is omitted throughout the paper and written,
for instance, as $\mu$ instead of $\mu I$.

\section{Solving the determining equations \label{soldet}}

In this section a solution for determining system of arbitrary dimension $n$ is presented.
Let us start with equation (\ref{qeq}).

Introducing new variable
\begin{gather}
\label{M}
 M=Q-\varphi,
\end{gather}
where scalar function $\varphi$ is defined in the following way
\begin{gather}
\label{phi}
\varphi=\left\{\begin{array}{ll}
 \lambda\tan(\lambda x+\gamma),& \nu=\lambda^2>0 \\[1 ex]
 -\lambda\tanh(\lambda x+\gamma),& \nu=-\lambda^2<0 \\[1 ex]
 -\frac1{x+\gamma},& \nu=0
\end{array}\right.
\end{gather}
and $\gamma\in\R$ is chosen so that $M$ is not a singular matrix, we can transform
equation (\ref{qeq}) into
\begin{gather}
\label{equM}
 M^{-1}M'M^{-1}=I+2\varphi M^{-1}.
\end{gather}

Note that
\begin{gather*}
 (M^{-1})'=-M^{-1}M'M^{-1}.
\end{gather*}

So if we set $N=M^{-1}$ then we obtain the following linear equation:
\begin{gather}
\label{equN}
 N'=-I-2\varphi N.
\end{gather}
Its general solution has the form
\begin{gather}
\label{solN}
 N=-\rho(x) I+\theta(x)C,
\end{gather}
where $\rho(x)$ and $\theta(x)$ are scalar real valued functions and $C$ is an arbitrary constant matrix,
\begin{gather}
\label{rhotheta}
\begin{split} &
\rho=\left\{\begin{array}{ll}
 \frac1{2\lambda}\sin(2(\lambda x+\gamma)),& \nu=\lambda^2>0 \\[1 ex]
 \frac1{2\lambda}\sinh(2(\lambda x+\gamma)),& \nu=-\lambda^2<0 \\[1 ex]
 x+\gamma,& \nu=0
\end{array}\right. \\ &
\theta=\left\{\begin{array}{ll}
 \cos^2(\lambda x+\gamma),& \nu=\lambda^2>0 \\[1 ex]
 \cosh^2(\lambda x+\gamma),& \nu=-\lambda^2<0 \\[1 ex]
 (x+\gamma)^2,& \nu=0
\end{array}\right.
\end{split}
\end{gather}

$Q$ is hermitian, so $N$ is hermitian too. Moreover, if unitary matrix
$U$ is such that $UNU^\dag $ is a diagonal matrix, then $UQU^\dag $ is also diagonal.
As $\rho$ and $\theta$ are scalar functions, then diagonalization transformation
$U$ should not depend on $x$. So $Q$ can be diagonalized
\begin{gather}
\label{diagQ}
 Q=\diag{q_1}{\dots,q_n}
\end{gather}
and equation (\ref{qeq}) transforms into system
\begin{gather}
\label{equqi}
 q_i'=q_i^2+\nu,\quad i=1... n,
\end{gather}
which has following solutions
\begin{gather}
\label{solqi}
\begin{split} &
 q_i=\lambda\tan(\lambda x+\gamma_i),\quad i=1...n,\quad \nu=\lambda^2>0; \\ &
 q_i=\left[\begin{array}{ll}
            -\lambda\tanh(\lambda x+\gamma_i),& i=1...m \\[1 ex]
	    -\lambda\coth(\lambda x+\gamma_i),& i=m+1...l \\[1 ex]
	    \pm\lambda,& i=l+1...n
           \end{array}\right.,\quad \nu=-\lambda^2<0; \\ &
 q_i=\left[\begin{array}{ll}
            -\frac1{x+\gamma_i},& i=1...m \\[1 ex]
	    0,& i=m+1...n
           \end{array}\right.,\quad \nu=0,
\end{split}
\end{gather}
where $\gamma_i\in\R,\quad i=1...n$ are integration constants.
In cases $\nu<0$ and $\nu=0$ matrix $Q$ consists of blocks of size
$m, l-m+1, n-l+1$ and $m, n-m+1$ correspondingly. Some of them may or may not appear,
i.e. have zero size.

So we define matrix $Q$ up to constant unitary transformation $U$.
Now let us consider linear equation (\ref{peq}) for $P=\{p_{ij}\}$ which can be solved element-wise:
\begin{itemize}
\item If $\nu=\lambda^2>0$
\end{itemize}
\begin{gather}
\label{solP1}
\begin{split} &
 p_{ii}=\frac{\mu}{\lambda}\tan(\lambda x+\gamma_i)+\varphi_{ii}\sec(\lambda x+\gamma_i),\quad i=1...n \\ &
 p_{ij}=\varphi_{ij}\sqrt{\sec(\lambda x+\gamma_i)\sec(\lambda x+\gamma_j)},\quad i=1...n, j=1...n
\end{split}
\end{gather}
\begin{itemize}
\item If $\nu=-\lambda^2<0$
\end{itemize}
\begin{gather}
\label{solP2}
\begin{split} &
 p_{ii}=\left[\begin{array}{ll}
  -\frac{\mu}{\lambda}\tanh(\lambda x+\gamma_i)+\varphi_{ii}\sech(\lambda x+\gamma_i),& i=1...m \\[1 ex]
  -\frac{\mu}{\lambda}\coth(\lambda x+\gamma_i)+\varphi_{ii}\csch(\lambda x+\gamma_i),& i=m+1...l \\[1 ex]
  \pm\frac{\mu}{\lambda}+\varphi_{ii}\exp(\pm\lambda x),& i=l+1...n
 \end{array}\right. \\ &
 p_{ij}=\left[\begin{array}{ll}
  \varphi_{ij}\sqrt{\sech(\lambda x+\gamma_i)\sech(\lambda x+\gamma_j)},& i=1...m, j=1...m \\[1 ex]
  \varphi_{ij}\sqrt{\sech(\lambda x+\gamma_i)\csch(\lambda x+\gamma_j)},& i=1...m, j=m+1...l \\[1 ex]
  \varphi_{ij}\sqrt{\sech(\lambda x+\gamma_i)\exp(\pm\lambda x)},& i=1...m, j=l+1...n \\[1 ex]
  \varphi_{ij}\sqrt{\csch(\lambda x+\gamma_i)\csch(\lambda x+\gamma_j)},& i=m+1...l, j=m+1...l \\[1 ex]
  \varphi_{ij}\sqrt{\csch(\lambda x+\gamma_i)\exp(\pm\lambda x)},& i=m+1...l, j=l+1...n \\[1 ex]
  \varphi_{ij}\exp(\pm\lambda x), \quad(\star)& i=l+1...n, j=l+1...n \\[1 ex]
  \varphi_{ij}, \quad(*)& i=l+1...n, j=l+1...n
 \end{array}\right.
\end{split}
\end{gather}
Use formula $(\star)$ if corresponding diagonal entries $q_i$ and $q_j$ have the same sign and formula $(*)$ otherwise.
\begin{itemize}
\item If $\nu=0$
\end{itemize}
\begin{gather}
\label{solP3}
\begin{split} &
 p_{ii}=\left[\begin{array}{ll}
              \frac{\varphi_{ii}-\frac{\mu x}2(x+2\gamma_i)}{x+\gamma_i},& i=1...m \\[1 ex]
	      -\mu x+\varphi_{ii},& i=m+1...n
              \end{array}\right. \\ &
 p_{ij}=\left[\begin{array}{ll}
              \frac{\varphi_{ij}}{\sqrt{(x+\gamma_i)(x+\gamma_j)}},& i=1...m, j=1...m \\[1 ex]
	      \frac{\varphi_{ij}}{\sqrt{x+\gamma_i}},& i=1...m, j=m+1...n \\[1 ex]
	      \varphi_{ij},& i=m+1...n, j=m+1...n
              \end{array}\right.
\end{split}
\end{gather}
where $\varphi_{ji}=\bar{\varphi_{ij}}\in\C$ are integration constants. Numbers $m$ and $l$ in the above
intervals correspond to the numbers defined in (\ref{solqi}). Matrix $P$ can be further
simplified with unitary transformation, that reduces the number of non-zero entries in similar blocks.
Obtained superpotentials are irreducible if there are enough non-zero entries $\varphi_{ji}$.

\section{Solving spectral problem \label{example}}

Consider the Schr\"odinger equation
\begin{equation}
\label{spectrprob}
 \hat{H}_k\psi=(H_k+c_k)\psi=E_k\psi,
\end{equation}
where $c_k$ vanishes with constant multiplied by unit matrix in the Hamiltonian $H_k$.
It follows from shape-invariant condition (\ref{shapeinv}) that
\begin{equation}
 \label{ck}
 C_k=c_{k+1}-c_k.
\end{equation}

Since all considered Hamiltonians are shape-invariant, equation (\ref{spectrprob}) can be solved
using the standard SSQM technique. An algorithm for constructing exact solutions of supersymmetric
shape-invariant Schr\"odinger equations can be found in \cite{Khare}.
\begin{itemize}
 \item Ground state $\psi_k^0(x)$ is proportional to the square-integrable solution of the first order equation
 \begin{equation}
  \label{grst}
  a_k\psi_k^0(x)\equiv\left(\frac\p{\p x}+W_k\right)\psi_k^0(x)=0.
 \end{equation}
 Function $\psi_k^0$ solves equation (\ref{spectrprob}) with eigenvalues
 \begin{equation}
  \label{grenergy}
  E_k^0=-c_k.
 \end{equation}

 \item Solution which corresponds to the $n^{th}$ excited state $\psi_k^n(x)$ can be represented as
 \begin{equation}
  \label{nthst}
  \psi_k^n(x)=a_k^\dag a_{k+1}^\dag\cdots a_{k+n-1}^\dag\psi_{k+n}^0(x)
 \end{equation}
 and must be a square-integrable function of $x$. The corresponding eigenvalue is
 \begin{equation}
  \label{nthenergy}
  E_k^n=-c_{k+n}.
 \end{equation}
 It follows from equations (\ref{ceq}), (\ref{ck}), (\ref{grenergy}), (\ref{nthenergy}) that
 \begin{equation}
  \label{nthenergycomplete}
  E_k^n=E_k^0 + 2n\mu - (n^2+2kn)\nu.
 \end{equation}
\end{itemize}
At the end of the section, an example of exactly integrable problem is presented.
Let $Q$ and $P$, defined by formulas (\ref{solqi}), (\ref{solP3}), be
\begin{equation}
 Q=\begin{pmatrix}
    -\frac1{x} && 0 \\
    0 && 0
   \end{pmatrix}, \quad
 P=\begin{pmatrix}
    \frac{\mu x}2-\frac1{2x} && -\frac\varphi{\sqrt{x}} \\
    -\frac\varphi{\sqrt{x}} && \mu x
   \end{pmatrix},
\end{equation}
where $\mu, \varphi$ are real constants, $\mu > 0$. Note that, in contrast with (\ref{solP3}),
the constant~$\mu$ is changed to $-\mu$ to simplify notation.
Then corresponding Schr\"odinger equation looks like
\begin{equation}
 -\frac{\p^2}{\p x^2}\vecphi
 +
 \begin{pmatrix}
  \frac{4k^2-1}{4x^2}+\frac{\mu^2x^2}4+\frac{\varphi^2}x-\mu k && \frac{\varphi k}{\sqrt{x^3}}-\frac{3\varphi\mu\sqrt{x}}2 \\
  \frac{\varphi k}{\sqrt{x^3}}-\frac{3\varphi\mu\sqrt{x}}2 && \mu^2x^2+\frac{\varphi^2}{x}
 \end{pmatrix}\vecphi = E_k\vecphi
\end{equation}
and $c_k$ is equal to $\mu$.

The square integrable ground state solution $\psi^0=\begin{pmatrix}\phi^0 \\ \xi^0\end{pmatrix}$ has the form
\begin{equation}
 \begin{array}{l}
  \phi^0=C_1\phi_1+C_2\phi_2, \\
  \xi^0=C_1\xi_1+C_2\xi_2
 \end{array}
\end{equation}
for non integer $k>0$, where
\begin{equation}
\label{gr1}
 \begin{array}{l}
  \phi_1 = x^{k+\frac12}e^{-\frac{\mu x^2}2}H_B(k, 0, -2-k, \frac{4\varphi^2}{\sqrt{\mu}},\frac{\sqrt{\mu}}2x), \\[1 ex]
  \xi_1 = \frac{\sqrt{\mu}}{2\varphi}x^{k+1}e^{-\frac{\mu x^2}2}H_B'(k, 0, -2-k, \frac{4\varphi^2}{\sqrt{\mu}},\frac{\sqrt{\mu}}2x)+ \\[1 ex]
  \frac\mu{2\varphi}x^{k+2}e^{-\frac{\mu x^2}2}H_B(k, 0, -2-k, \frac{4\varphi^2}{\sqrt{\mu}},\frac{\sqrt{\mu}}2x)
 \end{array}
\end{equation}
and
\begin{equation}
\label{gr2}
 \begin{array}{l}
  \phi_2 = x^{\frac12}e^{-\frac{\mu x^2}2}H_B(-k, 0, -2-k, \frac{4\varphi^2}{\sqrt{\mu}},\frac{\sqrt{\mu}}2x), \\[1 ex]
  \xi_2 = \frac{\sqrt{\mu}}{2\varphi}xe^{-\frac{\mu x^2}2}H_B'(-k, 0, -2-k, \frac{4\varphi^2}{\sqrt{\mu}},\frac{\sqrt{\mu}}2x)+ \\[1 ex]
  \frac{2k-\mu x^2}{2\varphi}e^{-\frac{\mu x^2}2}H_B(-k, 0, -2-k, \frac{4\varphi^2}{\sqrt{\mu}},\frac{\sqrt{\mu}}2x).
 \end{array}
\end{equation}
Here $H_B$ denotes the Heun Biconfluent function. For integer $k$ solutions (\ref{gr1}) and (\ref{gr2}) are
identical and second linearly independent solution of equation (\ref{grst}) should be used instead of (\ref{gr2}).
It is however cumbersome and is omitted in the paper.

Note that constants $C_1$ and $C_2$ should be chosen so, that
\begin{equation}
 \int_{-\infty}^{\infty}\!\!(\psi^0)^\dag\psi^0dx=1.
\end{equation}

The spectrum of the problem is described by formula
\begin{equation}
 E^n = -(2n+1)\mu.
\end{equation}

\section{Discussion \label{disc}}

In this section the possibility of transforming unknown function $F_k$ in the shape-invariance condition (\ref{hshapeinv})
to unit shift is discussed.

During the computation it was assumed that $F_k=k+1$, let us show it is a reasonable restriction.
Under invertible transformation of variables
\begin{gather}
k\to\alpha(k)
\end{gather}
the function $F_k=F(k)$ changes by a similar transformation
\begin{gather}
F(k)\to\alpha F\alpha^{-1}(k)=\alpha(F(\alpha^{-1}(k))).
\end{gather}
Searching for such transformation, that would change function $F(k)$ to unit shift, we get the equation
\begin{gather}
F(\alpha^{-1}(k))=\alpha^{-1}(k+1).
\end{gather}

The above equation is known as Abel functional equation. The results concerning the solution of this
equation were obtained in papers \cite{kork}, \cite{bel1}, \cite{bel2}, \cite{jit}.

Let $X$ be $\R$ or $\R^+$. It is proved that
\begin{itemize}
\item[(C1)] {\it if $F:X\to\R$ is an injective function 
such that for every compact set $K\subset X$
there exists $p\in\N$ such that $\forall n,m\in\N^0,\ \abs{n-m}\geq p:$
\begin{equation*}
F^n(K)\cap F^m(K)=\emptyset
\end{equation*}
then there exists a solution to the Abel functional equation.}
\end{itemize}

So if the above condition is fulfilled, $F_k$ can be transformed into unit shift.

\section{Conclusion \label{conc}}

Restricting the class of superpotentials to a simple form (\ref{superpot}),
it was possible to find new exactly solvable systems of Schr\"odinger equations. As corresponding
Hamiltonians satisfy shape-invariant condition (\ref{shapeinv}), the systems can be solved
using a standard SSQM technique. An illustrative example of such a system was presented.
Related superpotentials (\ref{solqi}) -- (\ref{solP3})
are matrices of arbitrary dimension $n$ that can be further simplified with unitary transformation.

The Hamiltonian factorization and the shape-invariance condition itself were discussed.
It was shown that an alternative representation (\ref{comrep})
of operators $a_k$ and $a_k^\dag$ is equivalent to a standard one (\ref{standrep}).
An unknown function $F_k$ that appears in form-invariance condition significantly relies on the
superpotential's form. If it equals to identical function $F_k=k$, the corresponding superpotential is
direct sum of shifted one-dimensional oscillators. In other cases, it is reasonable to consider
$F_k$ to be unit shift, as it was shown in the discussion section.

In spite of this, the class of obtained potentials is strongly restricted by the superpotential's form.

In future works, a more general form of superpotential will be discussed.

\subsection*{Acknowledgment}
The author thanks to Prof. Anatoly Nikitin for useful discussions and valuable comments.

\end{document}